\documentclass[utf8]{frontiersFPHY}
\setcitestyle{square}
\usepackage{url,hyperref,lineno,microtype,subcaption}
\usepackage[onehalfspacing]{setspace}
\usepackage{ads}

\newcommand{\dd}{\text{d}}
\newcommand{\e}{\text{e}}
\newcommand{\ii}{\text{i}}

\newcommand{\be}{\begin{equation}}
\newcommand{\ee}{\end{equation}}
\newcommand{\ben}{\begin{equation*}}
\newcommand{\een}{\end{equation*}}
\newcommand{\bea}{\begin{eqnarray}}
\newcommand{\eea}{\end{eqnarray}}
\newcommand{\bean}{\begin{eqnarray*}}
\newcommand{\eean}{\end{eqnarray*}}

\newcommand{\kpc}{\ensuremath{\,\mathrm{kpc}}}
\newcommand{\Mpc}{\ensuremath{\,\mathrm{Mpc}}}
\newcommand{\ev}{\ensuremath{\,\mathrm{eV}}}

\newcommand{\kms}{\ensuremath{\,\mathrm{km}\,\mathrm{s}^{-1}}}

\def\keyFont{\fontsize{8}{11}\helveticabold }
\def\firstAuthorLast{Zhang} 
\def\Authors{Jiajun Zhang$^{1,*}$,Hantao Liu$^{2}$,Ming-Chung Chu$^{2}$}
\def\Address{$^{1}$Department of Astronomy, School of Physics and Astronomy, Shanghai Jiao Tong University, No. 800 Dongchuan Road, 200240, Shanghai, China\\
$^2$Department of Physics, The Chinese University of Hong Kong, Hong Kong, China}
\def\corrAuthor{Jiajun Zhang}

\def\corrEmail{liamzhang@sjtu.edu.cn}

\begin{document}
\onecolumn
\firstpage{1}

\title[FDM Simulation]{Cosmological Simulation for Fuzzy Dark Matter Model} 

\author[\firstAuthorLast ]{\Authors} 
\address{} 
\correspondance{} 

\extraAuth{}

\maketitle

\begin{abstract}
Fuzzy Dark Matter (FDM), motivated by string theory, has recently become a hot candidate for dark matter. The rest mass of FDM is believed to be $\sim 10^{-22}\ev$ and the corresponding de-Broglie wave length is $\sim 1\kpc$. Therefore, the quantum effect of FDM plays an important role in structure formation. In order to study the cosmological structure formation in FDM model, several simulation techniques have been introduced. We review the current status and challenges in the cosmological simulation for the FDM model in this paper.     
\tiny
 \keyFont{ \section{Keywords:} Cosmology, Dark Matter, Simulation, Large Scale Structure, Halo, Fuzzy Dark Matter, Quantum Pressure}
\end{abstract}

\section{Introduction}

The nature of dark matter is one of the key mysteries of modern cosmology and physics. 
Dark matter is widely believed to be dominated by cold dark matter (CDM), 
supported by different observations such as the
mass-to-light ratio of clusters of galaxies~\citep{MLratio}, the
rotation curves of galaxies~\citep{einasto1974dynamical}, the Bullet
Cluster~\citep{Bullet}, the cosmic microwave background
(CMB)~\citep{Planck} and the large-scale structure of the
universe~\citep{MatterPowerSpectrum}. However, despite its success on
large scales, the CDM paradigm faces three problems on small scales, known as
the ``small-scale crisis''~\citep{SmallScaleCrisis}: (i) the missing
satellite problem, (ii) the cusp-core problem, and (iii) the too-big-to-fail 
problem. The key point of these problems is that CDM model predicts too much or too compact
structures on small scales. Two approaches are under discussion to solve these problems.
One is to smooth out the small-scale structure by astrophysical processes~\citep{AstroSolSSC},
and the other is to introduce alternative dark matter models 
like warm dark matter (WDM)~\citep{WDM}, decaying dark matter (DDM)~\citep{chen2015JCAP}, self-interacting dark
matter (SIDM)~\citep{SIDM} and fuzzy dark matter (FDM)~\citep{FDM}.

In the FDM model, the dark matter particles are made of ultra-light bosons in 
Bose-Einstein condensate (BEC) state \citep{ULAReview}. As an alternative to CDM, 
it suppresses small-scale structures while keeps the success of CDM on large scales
\citep{du2016substructure,Mocz:2017wlg,jjzhang2017lyman}. The FDM model is 
phenomenologically different from the CDM model due to its effective "quantum pressure" (QP) which originates from the uncertainty principle \citep{Hu:2000ke}. Apart from FDM, this 
model has many other names, such as wave dark matter ($\Phi$DM), ultra-light axion (ULA),
scalar field dark matter (SFDM), which is mainly due to historical reasons. 
These models have slightly different self-interactions and theoretical considerations. There are quite a few theoretical studies of such models. \citep{khlopov1985sfdm,sahni2000fcdm,chavanis2012bec,mishra2017alpha}.
However, models in which dark matter has no self-interaction are phenomenologically the same as FDM. 
The history of FDM and implementation is summarized in Ref.~\citep{Lee:2017qve}.

The predictions of FDM with mass $\sim 10^{-22}\ev$ are consistent with observations of 
the large-scale structure \citep{FDMCMBLSS}, high-$z$ galaxies, 
CMB optical depth~\citep{FDMHMF}, 
and the density profiles of dwarf spheroidal galaxies~\citep{FDMSim}. The tightest
constraints come from the comparison of the recent Lyman-alpha forest observations with
FDM hydrodynamic simulations \citep{irvsivc2017first, armengaud2017constraining, 2017arXiv170800015K}. These works claimed that FDM model with particle mass less than $10^{-21}\ev$ is ruled out at $95\%$ confidence level. 
However, it has been pointed out that the quantum pressure
plays quite non-trivial role in structure formation, which is neglected in the
hydrodynamic simulations for Lyman-alpha forest. The simulation uncertainties are also
important issues for making such tight constraints \citep{jjzhang2017lyman}.

In order to constrain the parameter space of the FDM model, or to look for smoking-guns for it, simulation is extremely important. There have been eight
different codes proposed to perform simulations for the FDM model \citep{FDMSim,mocz2015numerical,schwabe2016prd,veltmaat2016cosmological,Mocz:2017wlg,zhang2018ultra,nori2018ax,edwards2018py}. They can be classified into
two major approaches: solving the Schr\"odinger-Poission equation or the ``equivalent'' Madelung equations. We reviewed these works and summarized them into a table.
The pros and cons of these different simulation methods were clearly stated. We gave some comments
on the current status and challenges for FDM simulation.

The paper is organized in the following sections: we review the basic equations necessary for the FDM model in Sec.~\ref{Sec:Eq}, the simulation treatments and code comparison in Sec.~\ref{Sec:Sim}, the current status and challenges of FDM simulation in Sec.~\ref{Sec:Sum}, and finally we discuss about possible smoking-gun signatures for the FDM model.

\section{Basic Equations}\label{Sec:Eq}

To study the structures on galactic scales in the low red-shift universe, it is safe to ignore the self-interaction of the scalar field describing the FDM. 
The action has the following form
\begin{equation}
S=\int\dfrac{d^4 x}{\hbar c^2}\sqrt{-g}\left\{\dfrac{1}{2}g^{\mu\nu}\partial_{\mu}\phi\partial_{\nu}\phi-\dfrac{1}{2}\dfrac{m^2 c^2}{\hbar^2}\phi^2\right\},
\end{equation} 
where we follow the convention in Ref.\citep{Hui:2016ltb}.
The related de Broglie wavelength of particles with rest mass $m$ is
\begin{equation}\label{wavelength}
\dfrac{\lambda}{2\pi}=\dfrac{\hbar}{mv}=1.92\kpc\left(\dfrac{10^{-22}\ev}{m}\right)\left(\dfrac{10 \kms}{v}\right).
\end{equation}
Using the least action principle and WKB approximation in the non-relativistic limit, one can simplify the governing equations 
of the scalar field to the Schr{\"o}dinger-Poisson equations, 
\begin{equation}\label{schrodinger}
\ii\hbar\dfrac{\dd\Psi}{\dd t}=-\dfrac{\hbar^2}{2m}\boldsymbol{\nabla}^2\Psi+mV\Psi, 
\end{equation}
where $\Psi$ is the plane wave description of the scalar field $\phi$,
\begin{equation}
\phi=\sqrt{\dfrac{\hbar^3 c}{2m}}\left(\Psi \e^{-\ii mc^2t/\hbar}+\Psi^{*} \e^{\ii mc^2 t/\hbar}\right),
\end{equation}
and $V$ is  gravitational potential,
\begin{equation}\label{poisson}
\boldsymbol{\nabla}^2V=4\pi Gm|\Psi|^2. 
\end{equation}
The wave function $\Psi$ can be written as 
\begin{equation}\label{psi}
\Psi=\sqrt{\dfrac{\rho}{m}}\exp\left(\dfrac{\ii S}{\hbar}\right)
\end{equation}
in terms of the number density of FDM particles $\rho/m$, 
while we can define the gradient of $S$ to be the momentum,
\begin{equation}\label{velocity}
\boldsymbol{\nabla} S=m\boldsymbol{v}. 
\end{equation}
After transforming the wave function, the Schr{\"o}dinger-Poisson equations can be written in an equivalent fluid dynamics form with the continuity equation, 
\begin{equation}\label{continuity}
\dfrac{\dd\rho}{\dd t}+\boldsymbol{\nabla}\boldsymbol{\cdot}\left(\rho \boldsymbol{v}\right)=0, 
\end{equation}
and the Euler equation, 
\begin{equation}\label{momentum}
\dfrac{\dd\boldsymbol{v}}{\dd t}+\left(\boldsymbol{v}\boldsymbol{\cdot}\boldsymbol{\nabla}\right)\boldsymbol{v}=-\boldsymbol{\nabla}\left(Q+V\right), 
\end{equation}
where the quantum pressure $Q$ is defined as
\begin{equation}\label{pressure}
Q=-\dfrac{\hbar^2}{2m^2}\dfrac{\boldsymbol{\nabla}^2\sqrt{\rho}}{\sqrt{\rho}}=-\dfrac{\hbar^2}{2m^2}\left(\dfrac{\boldsymbol{\nabla}^2\rho}{2\rho}-\dfrac{\left\vert\nabla\rho\right\vert^2}{4\rho^2}\right).
\end{equation}
Eqs.~\eqref{continuity} and ~\eqref{momentum} are known as the Madelung equations
~\citep{spiegel1980fluid,Uhlemann:2014npa,Marsh:2015daa}. 
In cosmological simulations, we also need to consider the expansion of the universe.
Eq.~\ref{schrodinger} should be rewritten as:
\begin{equation}
\ii\hbar\left(\dfrac{\dd\Psi}{\dd t}+\dfrac{3}{2}H\Psi\right)=-\dfrac{\hbar^2}{2m}\boldsymbol{\nabla}^2\Psi+mV\Psi,
\end{equation}
where $H=\dot{a}/a$ is the Hubble parameter and $a$ is the scale factor of the universe.
The Madelung equations change accordingly
\begin{equation}\label{ConExp}
\dfrac{\dd\rho}{\dd t}+3H\rho+\dfrac{1}{a}\boldsymbol{\nabla}\boldsymbol{\cdot}(\rho \boldsymbol{v})=0, 
\end{equation} 
\begin{equation}\label{EurExp}
\dfrac{\dd\boldsymbol{v}}{\dd t}+H\boldsymbol{v}+\dfrac{1}{a}(\boldsymbol{v}\boldsymbol{\cdot}\boldsymbol{\nabla})\boldsymbol{v}=-\dfrac{1}{a}\boldsymbol{\nabla}V-\dfrac{1}{a^3}\boldsymbol{\nabla}Q.
\end{equation}
In short, the difference between the FDM and CDM models lies in the existence of the quantum pressure $Q$. 
The derivation of the quantum pressure is well described in text books about BEC. Since $Q\propto m^{-2}$, the quantum pressure in lab BEC systems is negligible. However, this effect is important in the FDM model whose particle mass is around $m\sim 10^{-22}\ev$.

From the linear perturbation of the equations \eqref{ConExp} and \eqref{EurExp}, the density contrast evolves according to  
\begin{equation}\label{linpert}
   \ddot{\delta}+2H\dot{\delta}+\left(\dfrac{\hbar^2k^4}{4m^2a^4}-\frac{4\pi G\bar{\rho}}{a^3}\right)\delta=0. 
\end{equation}
A solution is given by a plane wave with wave number
\begin{equation}
    k_J(a)=\left(\dfrac{16\pi G\bar{\rho}a^3m^2}{\hbar^2}\right)^{1/4}a^{1/4}.
\end{equation}
If $k<k_J(a)$, gravity dominates and the structure will collapse, while modes with $k>k_J(a)$ will expand due to the repulsive quantum pressure. So this is the Jeans wavenumber of the FDM model. The growing mode $D_+(k,a)$ and the decaying mode $D_-(k,a)$ of equation \eqref{linpert} are 
\begin{equation}
    \begin{array}{cc}
         D_+(k,a)&=\left[\left(3-x^2\right)\cos x+3x\sin x\right]/x^2  \\
         D_-(k,a)&=\left[\left(3-x^2\right)\cos x-3x\sin x\right]/x^2 
    \end{array},\quad x(k,a)=\sqrt{6}k^2/k_J^2(a).
\end{equation}
For $k\ll k_J(a)$, the two modes return to the CDM solutions $D_+\propto a$ and $D_-\propto a^{-2/3}$, meaning that the FDM and CDM have the exact same behavior on the large scales. On the other hand, for $k\gg k_J(a)$, the growth of the structure in FDM is suppressed because $D_+\propto k^{-4}$. But the Jeans wave number $k_J(a)\propto a^{1/4}$ is growing over time, and so the small-scale structures will eventually start growing: the smaller the scale (the larger the wave number), the later this mode started growing.

\section{Simulation Review}\label{Sec:Sim}

A typical N-body cosmological simulation contains the following steps:
\begin {enumerate}
\item Distribute the simulation particles in the simulation box homogeneously and isotropically. The FDM model has the same preparation of these pre-initial conditions as the CDM model.
\item Calculate the matter power spectrum at a relatively high redshift, such as $z=99$, according to the prediction of the linear perturbation theory. The modification brought by the FDM model can be either calculated by AxionCAMB\citep{axioncamb}, or given by the empirical transfer function Ref.\citep{Hu:2000ke},
\begin{equation}
P_{FDM}(k)=T_F^2(k)P_{CDM}(k),\quad T_F(k)\approx\dfrac{\cos x^3}{1+x^8},
\end{equation}
where $x=1.61\left(\dfrac{m}{10^{-22}\ev}\right)^{1/18}\dfrac{k}{k_{Jeq}},k_{Jeq}=9\left(\dfrac{m}{10^{-22}\ev}\right)^{1/2}\Mpc^{-1}$. It has been shown that using these two methods makes little difference, and the empirical transfer function is a good approximation\citep{armengaud2017constraining}.
\item Perturb the distribution of particles according to the matter power spectrum;
\item Solve the Euler equation and Poisson equation iteratively (continuity equation is naturally obeyed using N-body simulation) until the desired redshift, such as $z=0$.  To incorporate the quantum pressure into the Lagrangian particle tracking simulation scenario, there are four different codes available, summarized in Table \ref{tab.Lagrangian}.
\begin{table}
\centering
\caption{Summary of the Lagriangian based simulation codes for FDM model (Madelung Solvers).}\label{tab.Lagrangian}
\begin{tiny}
\begin{tabular}{ccccccc}
  \hline
  Author & Method (Code Base) & Cosmo-Sim & Granular structure & Solitonic Core & Activity & Open Source\\
  \hline
  Veltmaat  et al.\citep{veltmaat2016cosmological} & PIC (NyX) & Yes & No & Yes & Yes & No\\
  Mocz  et al.\citep{mocz2015numerical} & SPH & No & No& --  & No & No\\
  Nori  et al.\citep{nori2018ax} & SPH (P-Gadget3) & Yes & -- & -- & Yes & No\\
  Zhang  et al.\citep{zhang2018ultra} & PP (Gadget2) & Yes & No & Yes & Yes & Yes\\
  \hline
\end{tabular}
\end{tiny}
\end{table}
\end {enumerate}

The first three methods in Tab.\ref{tab.Lagrangian} are based on the traditional SPH method. The essence of the SPH method is to first assign all physical quantities (like density $\rho$, velocity $\boldsymbol{v}$ and pressure $P$) on each simulation particles, then calculate the physical fields by a special interpolation method --- kernel smoothing, which in turn give rise to the time evolution of the simulation particles through the Euler equation and the equation of state. The kernel smoothing of the field is simplly:
\begin{equation}
	O_i=\sum m_j\frac{O_j}{\rho_j}W\left(\dfrac{r_{ij}}{h}\right),
\end{equation}
where $W$ is a spherical function with finite support, $r_{ij}$ is the distance between two particles, and $h$ is the smoothing parameter, notice that it is not the dimensionless Hubble parameter. The quantum pressure \eqref{pressure}, however, not only depends on the field itself but also its derivatives up to second order. The three different implementations of the SPH methods listed above use three different method to calculate derivatives. 

Ref.~\citep{veltmaat2016cosmological} used the particle-in-cell method: 
\begin {enumerate}
    \item Assign the physical quantities of each simulation particle onto an auxiliary cubic grid;
    \item Calculate the derivatives of the physical fields with the finite difference method;
    \item Interpolate the derivatives of physical fields back to the positions of the simulation particles. 
\end {enumerate}
The additional force coming from quantum pressure is given by 
\begin{equation}\label{GridAcc}
    -\nabla Q_i=\dfrac{\hbar^2}{2m^2}\left(\Delta x\right)^3\sum_{j,k,l}\dfrac{\left(\nabla^2\sqrt{\rho}\right)_{j,k,l}}{\sqrt{\rho_{j,k,l}}}m_i\nabla W\left(\dfrac{\left\vert\boldsymbol{r}_i-\boldsymbol{x}_{j,k,l}\right\vert}{h}\right),
\end{equation}
where the seven-point stencil is used to calculate the Laplacian: $\left(\nabla^2\sqrt{\rho}\right)_{j,k,l}=\sqrt{\rho_{j+1,k,l}}+\sqrt{\rho_{j-1,k,l}}+\sqrt{\rho_{j,k+1,l}}+\sqrt{\rho_{j,k-1,l}}+\sqrt{\rho_{j,k,l+1}}+\sqrt{\rho_{j,k,l-1}}-6\sqrt{\rho_{j,k,l}}$.

Ref.~\citep{mocz2015numerical} and~\citep{nori2018ax} used similar formulae: 
\begin{equation}
    \nabla O_i=\sum m_j\dfrac{O_j-O_i}{\rho_j}\dfrac{\Theta_j}{\Theta_i}\nabla W\left(\dfrac{r_{ij}}{h}\right), \quad \nabla^2 O_i=\sum m_j\dfrac{O_j-O_i}{\rho_j}\dfrac{\Theta_j}{\Theta_i}\nabla^2 W\left(\dfrac{r_{ij}}{h}\right)-\dfrac{2}{\Theta_i}\nabla O_i\cdot\nabla\Theta_i,
\end{equation}
with different choices of the auxiliary function: $\Theta_i=1$ for Ref.~\citep{mocz2015numerical} and $\Theta_i=\sqrt{\rho_i}$ for Ref.~\citep{nori2018ax}. The force contributed by the quantum pressure is given by 
\begin{equation}\label{PartAcc}
    -\nabla Q_i=\dfrac{\hbar^2}{2m^2}\sum_j\dfrac{m_j}{f_j\rho_j}\left(\dfrac{\boldsymbol{\nabla}^2\rho_j}{2\rho_j}-\dfrac{\left\vert\nabla\rho_j\right\vert^2}{4\rho^2_j}\right)\nabla W\left(\dfrac{r_{ij}}{h}\right),
\end{equation}
where $f_j=1+\dfrac{h_j}{3\rho_j}\sum_k m_k \dfrac{\partial W\left(r_{jk}/h_j\right)}{\partial h_j}$ is a correcting factor when variable smoothing length is used~\citep{Springel2002}.

All the three methods above involve the estimation of density and its derivatives on the grids or at the positions of the particles, and the force \eqref{GridAcc} or \eqref{PartAcc} has the form of many body interaction. Therefore, their computational costs are relatively high. Ref.\citep{zhang2018ultra} improves the SPH method by reducing the quantum pressure to a two-body particle-particle interaction; hence the additional force can be easily added to the tree algorithm in the TreePM method without the need to resort to the SPH method, and the computational time are greatly reduced.

From Tab.\ref{tab.Lagrangian}, we conclude that all these Lagrangian based simulations cannot produce granular structures which are expected to appear as the result of quantum interference. There are two possible explanations that may be viewed as the fundamental flaws of Lagrangian based simulations of the FDM model:
\begin{itemize}
\item The Schr{\"o}dinger-Poisson equations and Madelung equations are not strictly equivalent. As proved in Ref.~\citep{wallstrom1994pra}, a quantization condition $\oint_L\boldsymbol{v}\cdot\dd\boldsymbol{l}=2\pi j$ ($j\in \mathbb{Z}$ and $L$ is any closed loop.) is necessary to recover the Schr{\"o}dinger-Poisson equations from the Madelung equations, which is not checked and possibly not obeyed in Lagrangian based simulations.
\item The smoothing kernel method which is indispensable in Lagrangian based simulations cannot accurately estimate the matter density field and its second order derivative simultaneously if merely a single smoothing length is used. As proved in Ref.~\citep{Silverman1978Biometrika}, the relative error of the estimation of the second order derivative of the density field could be as large as 100\% when the smoothing length is chosen to minimize the error of density estimation\citep{Merritt1996, Dehnen2001}, to solve the Poisson equation.
\end{itemize}

It is a consensus that in the center of a virialized FDM halo, there is a solitonic core made of wave function in the ground state with the same phase\citep{schive2014understanding}. Although the core-like structures appear in the Lagrangian based simulations, they are not trustworthy due to the two reasons listed above.

Apart from Lagrangian based simulations, FDM model can also be studied by Eulerian based simulations summarized in Table \ref{tab.Eulerain}. The physical fields on the grid also need to be suitably set at the initial moment according to the cosmological linear theory prediction. The time evolution of the wave function is given by 
\begin{equation}
    \Psi\left(\boldsymbol{x},t+\Delta t \right)=T\exp\left[-\dfrac{\text{i}\Delta t}{\hbar}\int\text{d}t'\left(-\dfrac{\hbar^2}{2m}\nabla^2+mV\left(\boldsymbol{x},t'\right)\right)\right]\Psi\left(\boldsymbol{x},t\right)
\end{equation}
where $T$ is the time-ordering symbol. For a sufficiently small time step, it can be approximated as
\begin{equation}
    \Psi\left(\boldsymbol{x},t+\Delta t \right)=\exp\left(\dfrac{\text{i}\hbar\Delta t}{2m}\nabla^2-\dfrac{\text{i}m\Delta t}{2\hbar}V\left(\boldsymbol{x},t+\Delta t\right)-\dfrac{\text{i}m\Delta t}{2\hbar}V\left(\boldsymbol{x},t\right)\right)\Psi\left(\boldsymbol{x},t\right),
\end{equation}
which can be further splitted into three operations according to the Baker–Campbell–Hausdorff formula:
\begin{equation}
    \Psi\left(\boldsymbol{x},t+\Delta t \right)=\exp\left(-\dfrac{\text{i}m\Delta t}{2\hbar}V\left(\boldsymbol{x},t+\Delta t\right)\right)\exp\left(\dfrac{\text{i}\hbar\Delta t}{2m}\nabla^2\right)\exp\left(-\dfrac{\text{i}m\Delta t}{2\hbar}V\left(\boldsymbol{x},t\right)\right)\Psi\left(\boldsymbol{x},t\right).
\end{equation}
This formula has a close resemblance to the kick-drift-kick time evolution in the particle method. The "kick" step is done in real space, which effectively just changes the phase angle at each point. The "drift" step is completed in the Fourier space:
\begin{align}
     D\left(\dfrac{\Delta t}{2}\right)\Psi\left(\boldsymbol{x},t\right)&=\text{IFFT}\left\{ -\dfrac{\text{i}\hbar }{m}\dfrac{\Delta t}{2}k^2 \text{FFT}\left[\Psi\left(\boldsymbol{x},t\right)\right] \right\}.
\end{align}

The main differences between the Eulerian based and Lagrangian based methods are that the original Schr{\"o}dinger-Poisson equations are solved in the former, not the transformed Madelung equations as in the latter, and the Eulerian method can be used to reliably estimate second order derivatives of the fields. From Tab.~\ref{tab.Eulerain}, we find that most of the Eulerian based simulations can produce granular structures and solitonic cores. The sizes of the simulation boxes in these simulations, however, are not large enough to be considered as cosmological scale simulations. The daunting computational costs make it too difficult to perform simulations with box size larger than $10\Mpc/h$

\begin{table}
\centering
\caption{Summary of the Eulerain based simulation codes for FDM model (Schr{\"o}dinger-Poisson Solvers).}\label{tab.Eulerain}
\begin{tiny}
\begin{tabular}{ccccccc}
  \hline
  Author & Method (Code Base) & Cosmo-Sim & Granular structure & Solitonic Core & Activity & Open Source\\
  \hline
  Schive et al.\citep{schive2014understanding} & AMR (GAMER) & No & Yes & Yes & Yes & No\\
  Schwabe  et al.\citep{schwabe2016simulations} & AMR (Nyx) & No & Yes & Yes & Yes & No\\
  Mocz  et al.\citep{Mocz:2017wlg} & Moving-mesh (AREPO) & No & Yes & Yes & Yes & No\\
  Edwards  et al.\citep{edwards2018py} & Grid & No & Yes & -- & Yes & Yes\\
  \hline
\end{tabular}
\end{tiny}
\end{table}

For current simulation codes, there have to be a trade-off between the fidelity of the simulations and the scale of the simulations. On one hand, both the granular structures and the solitonic cores are smoking gun features of the FDM model, and it is very important to understand their properties. On the other hand, cosmological scale simulations are needed  whenever large scale survey data are used to constrain the FDM model or the properties of galaxy cluster are studied.

To simplify the generation of the granular structures, a self-consistent method was introduced in Ref.~\citep{lin2018self}. In their simplified model, the halo is composed of "smooth" density distribution along the radial direction and "granular" interference structure along the angular direction, the radial direction density profile is given by a typical guess and the angular direction density distribution is described by spherical harmonics. By fitting to the simulations, they find that the fermionic King model is the best fit energy distribution function and the generated halo is quite similar to that in simulation. With this method, they can generate a halo as massive as Milky Way ($\sim10^{12}M_{\odot}$) with granular structures.  However, it is still very difficult to self-consistently construct a very massive halo, such as a cluster scale halo ($\sim10^{14}M_{\odot}$), because of the limited computational power and poor generating algorithm. Making use of the information about the granular structures, a very promising smoking gun detection method was introduced in Ref.~\citep{martino2017prl,khmelnitsky2014pta,porayko2014pta} with Pulsar Timing Array. By studying the modulation of the arriving time of pulses from many pulsars, the Pulsar Timing Array can be used to directly detect the granularity of dark matter distribution in the Milky Way galaxy. The Parkes Pulsar Timing Array collaboration has obtained the first constraints of FDM as $m>10^{-23}\ev$ \citep{PPTA2018FDM}. This method can not only set constraints on the rest mass of FDM particles, but also confidently claim the existence of FDM. Only if we understand the granular structures of the Milky Way halo in much detail, we can make correct predictions for the modulation pattern observed in Pulsar Timing Array.

The other smoking gun feature of the FDM model is the existence of solitonic cores in the dark matter halos. Different simulation groups reported that they have found solitonic cores \citep{FDMSim,schive2014understanding,schwabe2016prd,Mocz:2017wlg}. The solitonic core solution can also be obtained analytically \citep{chavanis2011analytical,chavanis2011numerical}. The numerical simulations \citep{FDMSim,schive2014understanding} provide the empirical core-halo mass relation,
\begin{equation}
x_c \approx 160(\dfrac{M_h}{10^{12}M_{\odot}})^{-\dfrac{1}{3}}(\dfrac{m}{10^{-22}\ev})^{-1}\mathrm{pc},\\\
\rho(x) \approx \dfrac{190 (\dfrac{m}{10^{-22}\ev})^{-2}(\dfrac{x_c}{100\mathrm{pc}})^{-4}}{(1+0.091(\dfrac{x}{x_c})^2)^8}\mathrm{M_\odot}\mathrm{pc}^{-3},
\end{equation} 
where $x_c$ is the solitonic core radius, $m$ is the FDM particle mass and $M_h$ is the halo mass. The radius of the solitonic core is defined as the radius where the mass density drops by a factor of 2 from its value at the origin. This relation can be used to compare with observations. It is believed that the dark matter halos made of FDM should host a solitonic core in the center and follow usual NFW profile in the outer region. However, how the density profile transfers from soliton to the outer NFW profile is not consistent among the simulations \citep{bar2018rc}. It was claimed that the central solitonic core profile with outer NFW profile provide better fits than CDM predictions for the dwarf galaxies (especially for Fornax) \citep{FDMSim}. However, it faces challenge from the analysis of rotation curves of many other galaxies \citep{bar2018rc}.

FDM simulations with the largest box size ($50\Mpc/h$) were performed in Ref.~\citep{jjzhang2017lyman}. The limit of such simulations is their poor resolution, and they are unable to resolve the granular structures. The advantages of the Madelung solvers are their efficiency and mature related data analysis tools. Measured by the two large simulations Ref.~\citep{jjzhang2017lyman} and Ref.~\citep{nori2018ax}, whose algorithms are different, the matter power spectrum at $z=0$ is not only suppressed at small scales by the modification of the initial conditions, but also by the effect of quantum pressure with an additional $\sim10\%$ suppression at small scales. Such suppression is well expected and confirmed by different simulations. However, none of the Schr{\"o}dinger-Poisson solvers measured the matter power spectrum due to their small box sizes. Therefore, it is still under investigation about the effect of quantum pressure in the structure formation. It is still inconclusive from simulations how much suppression of the matter power spectrum can be introduced by the quantum pressure. In order to study large scale structure and use observations like weak lensing and red shift distortion to constrain FDM model, a much larger box size ($\sim500\Mpc/h$) is necessary. This is still a challenge for all the codes, among which the Madelung solvers are more hopeful to reach such a goal.

Recently, it was claimed in Ref.~\citep{2018arXiv180708153D} that the dynamical evidence of the existence of the solitonic core in the center of Milky Way was found. It was also claimed in Ref.~\citep{schive2014understanding} that the existence of the solitonic core can solve the cusp-core problem. Both of these two studies favor a FDM with $m\sim10^{-22}\ev$. However, the recent constraints from Lyman-alpha forest implied that FDM with $m<10^{-21}\ev$ was ruled out\citep{armengaud2017constraining,irvsivc2017first}. Such a tension is a problem for FDM models. Ref.~\citep{jjzhang2017lyman} suggested that considering of the important role of quantum pressure and systematic uncertainties may relieve this tension, but Ref.~\citep{nori2018lyman} with further analysis concluded that the quantum pressure can not affect the lyman-alpha forest significantly. Ref.~\citep{Hui:2016ltb} suggested that other astrophysical processes like patchy reionization can relieve the tension. Therefore, FDM with $m\sim10^{-22}\ev$ is still not conclusively ruled out by the observations of Lyman-alpha forest, while more serious studies are clearly needed. 

Other than the constraints from Lyman-alpha forest, the thickness of the stellar stream and the recent EDGES experiment also set constraints on FDM with $m>5\times10^{-21}\ev$ and $m>1.5\times10^{-22}\ev$\citep{2018arXiv180800464A,lidz2018prd}, respectively. A recent study of rotation curves of near-by galaxies also claimed that FDM with $m<10^{-21}\ev$ is not favored \citep{bar2018rc}. These independent constraints using different methods are also not supporting the value $m\sim10^{-22}\ev$ needed to solve the small-scale crisis. If more arguments and modifications to the current FDM model are made, we can relieve the tensions but lose the beauty and simplicity of the FDM model. 

We notice that with $m>10^{-22}\ev$, FDM model may still be able to solve the small-scale crisis due to the runaway tidal disruption. The runaway tidal disruption was found in simulations that the solitonic cores in the center of halos can be easily tidal disrupted in a runaway pattern when they rotate around the central massive halo \citep{du2018tidal}. This effect has not been studied by other codes yet, whether it is a physical effect or numerical illusion is unknown. A systematic study however, considering such mechanism needs simulations with sufficiently large box size ($>100\Mpc/h$) and high resolution ($<10^9 M_\odot/h$) at the same time. This is very difficult to reach now.
      
\section{Summary}\label{Sec:Sum}
In this paper, we have reviewed the basic idea of Fuzzy Dark Matter (FDM) model and the current status of simulations for this model. As a mini review, this work provides a short summary for the readers to follow the state-of-the-art research of FDM.

The cosmological simulation is important for understanding the structure formation and looking for smoking-gun signatures for the FDM model. The current simulation codes are not adequate to study the large-scale structure and halo properties under the framework of FDM. The codes designed to solve the SP equations are highly accurate, and many important features of the FDM model such as solitonic cores and granular structures are discovered. But these codes are too computationally heavy to perform simulations with large box size. The codes designed to solve the Madelung equations are less accurate, but the suppression of the matter power spectrum and halo mass function are given by simulations with these codes. These codes are more efficient and compatible with existing data analysis tools, but not accurate enough to resolve granular structures. All of the current FDM cosmological simulations are not large enough in terms of the box size to study large-scale structures systematically. Much efforts are needed to improve these methods.

The FDM model is an interesting alternative to the CDM model. The "small-scale crisis" in CDM might be solved in the FDM model, but more studies are needed to confirm this suggestion. The tensions from different observations on the rest mass of FDM particles may be relieved in many ways, such as considering the systematic uncertainties in the simulations, invoking astrophysical processes and new mechanisms in the FDM model. In order to understand the structure formation under the framework of FDM model, more and better simulations are important.

With the Madelung solvers, the simulations with box size $\sim500\Mpc/h$ can be expected in the near future, which is sufficient to constrain FDM models with observations such as weak lensing and red shift distortion. We need to first make sure that all different codes draw to the same conclusion about these observations. We also need to make sure that all the observable we measured both in observations and in simulations are consistent with the FDM framework. The effect of quantum pressure on the large scale structures can be degenerate with other models such as Warm Dark Matter (WDM), Self-Interacting Dark Matter (SIDM), Decaying Dark Matter (DDM) and so on. Therefore, even if we find conclusive evidence that the small scale structures are suppressed from observations, it is still not conclusive to claim that the FDM model is the correct model of the dark matter. On the other hand, the Schr{\"o}dinger-Poisson Solvers disclose the possibility of looking for smoking gun signals of the FDM model. Both the existence of a solitonic core in the center of a dark matter halo and the granular structure of dark matter halos are unique features of the FDM model, different from all the other models. It is possible to find granular structures for the the FDM model, using Pulsar Timing Array. It is also possible to rule out the FDM model by the next generation observations and more careful data analysis. The FDM model is a beautiful model with no more free parameters than the CDM model together with the WIMP assumption. If we can determine the mass range of FDM particles, it will significantly improve our understanding of dark matter and the universe.

\section{Acknowledgement}
J. Z acknowledges the support from China Postdoctoral Science Foundation 2018M632097.
\bibliographystyle{apsrev4-1}
\bibliography{MDM}

\end{document}